\documentclass{article}
\usepackage{fullpage, graphicx, amsmath,hyperref} 
\usepackage[english]{babel}
\usepackage[style=nature]{biblatex}
\title{What makes a study design quasi-experimental? The case of difference-in-differences}
\author{Audrey Renson$^1$, Daniel Westreich$^2$\\
\\
\small $^1$Department of Population Health, New York University Grossman School of Medicine, New York, United States\\%
\small $^2$Department of Epidemiology, Gillings School of Global Public Health, University of North Carolina at Chapel Hill\\%
}
\date{
    \today
}
\date{August 2025}

\begin{document}

\maketitle

\begin{abstract}
Study designs classified as quasi- or natural experiments are typically accorded more face validity than observational study designs more broadly. However, there is ambiguity in the literature about what qualifies as a quasi-experiment. Here, we attempt to resolve this ambiguity by distinguishing two different ways of defining this term. One definition is based on identifying assumptions being uncontroversial, and the other is based on the ability to account for unobserved sources of confounding (under assumptions). We argue that only the former deserves an additional measure of credibility for reasons of design. We use the difference-in-differences approach to illustrate our discussion.
\end{abstract}
\section{Introduction}
Study designs classified as quasi- or natural experiments are typically accorded more face validity than observational study designs more broadly. However, there is ambiguity in the literature about what qualifies as a quasi-experiment. Here, we attempt to resolve this ambiguity by distinguishing two different ways of defining this term, arguing that only one deserves an additional measure of credibility for reasons of design. We use the difference-in-differences approach to illustrate our discussion. 

\section{Quasi-experiments: two definitions}
Though some authors make a distinction between quasi- and natural experimental designs \cite{devocht2021conceptualising}, these terms are interchangeable for our purposes and we will use “quasi-experimental” going forward.

To the extent that it is defined, most definitions (e.g. \cite{devocht2021conceptualising,reeves2017checklist,caniglia2020did,dunning2008improving,porta2016dictionary,barnighausen2017introduction}) are similar to one provided by Greenland \cite{greenland2017foragainst}, which we paraphrase as follows: a quasi-experimental design is an observational study design in which a set of uncontroversial assumptions makes a causal effect identified (definition 1). Typically, these uncontroversial assumptions amount to an exposure, or a particular cause of the exposure, being independent of potential outcomes \cite{devocht2021conceptualising,reeves2017checklist,caniglia2020did,dunning2008improving,porta2016dictionary,barnighausen2017introduction}. One reason for this is that such independencies can often seem uncontroversial – and sometimes verifiable – based on logic or intuition. For example, the randomness of lotteries makes it uncontroversial that winning a lottery is independent of the health outcomes a lottery participant would have had, had they not won. On the other hand, functional form assumptions, such as a parametric regression model being correctly specified, are usually far more difficult to justify \textit{a priori}. 

A different definition that also appears in the literature considers quasi-experimental designs to be those that can ``adjust for unobserved confounding'' (definition 2) \cite{reeves2017checklist,khullar2021natural}. At first glance, definitions 1 and 2 may seem to overlap. When unmeasured confounding occurs for one group (say the general population) but not another (such as those individuals close to a cutoff in a regression discontinuity, or whose exposure is changed by an instrumental variable [IV]), by shifting attention to the latter group, we have in essence accounted for unmeasured confounding. 

However, on more careful examination, we see space between the two definitions.  Methods that adjust (or otherwise account) for unmeasured confounding generally do so under assumptions, which may or may not be uncontroversial. For example, quantitative bias analysis accounts for unmeasured confounding by assuming a known data-generating mechanism, and could therefore meet definition 2. However, quantitative bias analysis would not typically meet definition 1, because the assumption that the specified data generating mechanism is correct is rarely uncontroversial. On the other hand, IV methods can meet both definitions, but for different reasons. Though IV methods can be said to adjust for unmeasured confounding (thus meeting definition 2), it is the credibility of the IV assumptions in a particular setting that determines their meeting definition 1.
The question now arises: are both these definitions appropriate? Are both legitimate ways of identifying quasi-experimental studies, with all the additional “bonus” face-validity gained by being designated as quasi-experimental brings in public perception? We do not think so: we think that definition 2, in particular, should be avoided. 

\section{The case of difference-in-differences}

By way of illustration, we consider whether difference-in-differences (DID) should be considered a quasi-experimental design, as is common \cite{khullar2021natural,dore2024methods, westreich2019epidemiology, meyer1995natural}. We argue that DID does not in general meet definition 1, but may---under strong assumptions---meet definition 2. Therefore, we argue it should not in general be considered a quasi-experimental design.

We define DID as follows. Suppose we wish to estimate the effect of a binary treatment $A$ (e.g., a policy) that is newly adopted at a fixed time for some members of the population. Let $Y_{0i}$ denote an outcome measured at time $t=0$ in participant $i$ before anyone adopted treatment, let $Y_{1i}$ denote the outcome measured at time t=1 in participant $i$ after some individuals adopted treatment, and let $Y_{ti}^a$ denote a potential outcome for participant $i$ at time $t$ $(t=0,1)$ had $A$ been set by intervention to $a$. (We hereafter suppress $i$ subscripts unless needed to resolve ambiguity.) DID relies on the following three unverifiable assumptions: (i) causal consistency ($A_i=a$ implies $Y_{i1}=Y_{i1}^a$), (ii) no anticipation ($Y_0=Y_0^a$ for all $a$, meaning that outcomes prior to the treatment are not influenced by the treatment assignment), and (iii) parallel trends ($E[Y_1^0-Y_0^0|A=1]=E[Y_1^0-Y_0^0|A=0]$, meaning the untreated potential outcomes follow a constant average trend across treatment groups) \cite{callaway2022did}. Under these three assumptions, the average effect of the treatment in the treated (ATT) is identified as
\begin{equation}\label{eq:id}
    E[Y_1^1-Y_1^0|A=1]=E[Y_1-Y_0|A=1]-E[Y_1-Y_0 |A=0],
\end{equation}
a difference in mean differences (hence the name) which is straightforward to estimate via sample moments or a saturated regression (the steps to arrive at this identification result are provided in the \ref{app}). Though we have only described the simplest form of DID, the argument we make below also applies to many extensions of DID (e.g. parallel trends conditional on covariates \cite{abadie2005semiparametric,lechner2011did}, or DID with variation in treatment timing \cite{callaway2022did}).
To see that DID does not align with definition 1, consider first that assumptions i-iii make no reference to a treatment assignment mechanism, random or otherwise. This is in concert with other formal expositions of DID \cite{callaway2022did,abadie2005semiparametric,lechner2011did}. Instead of assuming that treatments are independent of potential outcomes, DID relies on parallel trends.
Parallel trends is a functional form assumption, similar to assuming a correctly specified parametric model. To see this, note that the concept of a ``trend'' is scale-dependent: we expressed parallel trends on the additive scale, but this is an arbitrary choice.  For example, suppose we had instead specified parallel trends on the ratio scale: $\frac{E[Y_1^0|A=1]}{E[Y_0^0|A=1]}=\frac{E[Y_1^0|A=0]}{E[Y_0^0|A=0]}$. Then, instead of (\ref{eq:id}), our identifying formula for the ATT would be $E[Y_1|A=1]-E[Y_0|A=1]\times\frac{E[Y_1|A=0]}{E[Y_0|A=0]}$. Other scales are equally possible (for example, additive in the mean of the log-transformed outcomes), and lead to identification \cite{roth2023parallel}.

Scale-dependence makes it difficult for parallel trends to be uncontroversial. One reason for this is that parallel trends on one scale typically implies non-parallel trends on another scale, for the same reason that lack of effect modification on one scale typically implies modification on at least one other scale \cite{westreich2019epidemiology}. Though it might appear plausible that two groups would experience equal changes over time, people might disagree on which scale these ``changes'' should be equal, and there is typically no logical reason to prefer one scale. Troublingly, in most applications, we will get a different effect estimate depending on the scale on which parallel trends is assumed \cite{meyer1995natural,roth2023parallel}. 

Next, we will show how DID can, under strong assumptions, align with definition 2. First, note that it is possible for parallel trends to hold in the presence of unobserved confounding. For example, suppose for a binary outcome we had $E[Y_1^0|A=0]=0.1$ and $E[Y_1^0|A=1]=0.4$; i.e., a treated group with four times the counterfactual risk of the untreated group, indicating strong confounding.  Parallel trends would still hold if the two groups had followed the same additive trend over time (e.g., if $E[Y_0^0|A=0]=0.05$ and $E[Y_0^0|A=1]=0.35$, so that the trend was $-0.05$ in both groups).  

To see how DID may still work in the presence of such unobserved confounding, take the DID formula for the ATT (repeated from eq. (\ref{eq:id}))
\[
    E[Y_1-Y_0|A=1]-E[Y_1-Y_0|A=0]
\]
and rearrange it as follows \cite{sofer2016negative}: 
\[
(E[Y_1|A=1]-E[Y_1|A=0])-(E[Y_0|A=1]-E[Y_0|A=0])
\]
In this form, we can see that what DID does is to take the simple post-treatment comparison, and subtract off the term $E[Y_0|A=1]-E[Y_0|A=0]$, which we will denote $c_0$. This term is the difference in outcome means before treatment was adopted for both groups, and is thus a measure of the baseline differences. Suppose that instead of DID, we attempted to estimate the ATT using the post-treatment comparison
\[
E[Y_1|A=1]-E[Y_1 |A=0]
\]
This comparison would be biased if there were any confounding; confounding would be represented as the difference between this estimate and the true ATT (expressed in potential outcomes). This difference is:
\begin{align*}
    \underbrace{\{E[Y_1|A=1]-E[Y_1|A=0]\}}_{\text{naïve post-treatment comparison}}&-\underbrace{\{E[Y_1-Y_1^0|A=1]\}}_{\text{true ATT}}\\
    ={E[Y_1|A=1]-E[Y_1|A=0]}&-{E[Y_1|A=1]-E[Y_1^0|A=1]}\\
=E[Y_1^0|A=1]-E[Y_1|A=0]&=c_1
\end{align*}
This last term ($c_1$) is the difference between the average untreated potential outcome in the treated group (which is unobservable) and the average outcome in the untreated group. Thus, the true ATT can be written as $E[Y_1|A=1]-E[Y_1|A=0]-c_1$; i.e., the post-treatment comparison adjusted for the bias of this comparison.

Now we can consider definition 2. In order to ``adjust for unobservable sources of confounding'' \cite{reeves2017checklist}, DID requires us to assume that the bias due to confounding, $c_1$, equals a specific value: the baseline difference $c_0$. In fact, the assumption that $c_1=c_0$ is simply a restatement of the parallel trends assumption \cite{lechner2011did,sofer2016negative}. Therefore, one way to understand the scale dependence of parallel trends is as an assumption of a constant magnitude of confounding bias over time. We assumed this constancy on the additive scale, but it could easily have been specified in terms of a constant risk ratio, odds ratio, or other measure. For the same reason that lack of effect modification on one scale typically implies modification on at least one other scale \cite{westreich2019epidemiology}, confounding bias can usually not be constant on multiple scales, and therefore DID is typically sensitive to this choice. 

\section{Discussion}

The notion of stolen valor in the military arises when someone lies about military service, such as wearing a medal they did not earn, for material gain and/or in a way that might dilute the meaning of such an honor \cite{afba2024stolen}. To call a study quasi-experimental is a kind of methodological medal, which accords that study the honor of greater face validity for causal interpretations; when such greater face validity is unearned or unwarranted, we can think of that designation as a sort of stolen methodological valor. It is dangerous, both because it may lead people to trust the study more than they should, and because it dilutes the label of ``quasi-experimental'' more broadly.

The credibility afforded to quasi-experiments is warranted for definition 1, as in this case credibility is granted in proportion to the degree to which the assumptions are uncontroversial.  This credibility is not warranted in general for definition 2: a method being able to adjust for unmeasured confounding under assumptions is not, by itself, a reason for credibility. It is instead the plausibility of the assumptions under which the adjustment occurs that should drive credibility. We therefore argue that definition 2 is not adequate to classify a method as ``quasi-experimental,'' and further that using the descriptor ``quasi-experimental'' for studies that only meet definition 2 constitutes stolen methodological valor, as described above.

Since DID does not align with definition 1, we would argue that it is important for epidemiologists not to afford DID any more credibility a priori than many other epidemiologic analysis methods; one way to help ensure this is to avoid stolen valor by not referring to DID studies as quasi-experimental. 

One possible reason that DID is grouped with quasi-experiments is that methodological guidance often recommends using DID primarily for settings with plausibly random treatments \cite{caniglia2020did,doleac2019evidence}, and DID is sometimes used in such cases \cite{novak2017immigration}. This would seem to conflict with our assertion that DID does not align with definition 1. Indeed, parallel trends (on any scale) will hold if the exposure is randomly assigned \cite{meyer1995natural,roth2023parallel}. However, a plausibly random treatment renders DID unnecessary (and usually inefficient \cite{yang2001efficiency}); a simple post-treatment comparison will suffice and is preferable \cite{dukes2025change}. 

Instead, prominent examples suggest that DID is typically used in cases where the treatment is not plausibly random. Consider one of the most commonly cited examples: John Snow’s 1856 study of the effects of water source on cholera mortality \cite{snow1856cholera}. In this case, there were substantial observed differences in the mean outcome between the treated and untreated groups prior to the treatment: cholera death rates of 14 vs. 8 per 1,000, suggesting that treatment assignment may not have been as good as random.  More broadly, DID is often applied in state-level comparisons of policies, where the underlying differences between states are often thought to be large \cite{matthay2022causal}, suggesting that the ability to adjust for unmeasured confounding under (unverifiable) assumptions is often key in the choice to use DID.

Instead of considering DID a quasi-experimental design, it may be more useful to think of DID as a method to analyze observational longitudinal data that makes different (untestable) assumptions than other methods (e.g., regression analysis, g-computation, inverse probability weighting). Accordingly, we should expect that investigators defend the assumptions behind applications of DID on substantive grounds using similar means as are otherwise recommended in observational studies; for example, using causal diagrams \cite{renson2025parallel}, rather than allowing a quasi-experimental label---one which is in our opinion frequently undeserved---do that work for them.

\printbibliography

\appendix
\renewcommand{\thesection}{Appendix}
\section{}\label{app}
Here we show how assumptions (i)-(iii) can be used to equate the ATT to the difference in observed outcome mean differences (\ref{eq:id}). Starting from the ATT we first use the fact that expectations are linear (i.e., $E[Y+X]=E[Y]+E[X]$):
\[
E[Y_1^1-Y_1^0|A=1]=E[Y_1^1|A=1]-E[Y_1^0|A=1]
\]
Adding and subtracting $E[Y_0^0|A=1]$, we have:
\[
=E[Y_1^1|A=1]-E[Y_0^0|A=1]-E[Y_1^0|A=1]+E[Y_0^0|A=1]
\]
Again using the fact that expectations are linear:
\[
=E[Y_1^1-Y_0^0|A=1]-E[Y_1^0-Y_0^0|A=1]
\]
Applying parallel trends, we can replace untreated potential outcome trends in the treated with those in the untreated:
\[
=E[Y_1^1-Y_0^0|A=1]-E[Y_1^0-Y_0^0|A=0]
\]
Applying causal consistency, we can replace $Y_1^1$ with $Y_1$ among the treated, and $Y_1^0$ with $Y_1$ among the untreated:
\[
=E[Y_1-Y_0^0|A=1]-E[Y_1-Y_0^0 |A=0]
\]
Applying no anticipation, we can replace $Y_0^0$ with $Y_0$, arriving at equation (\ref{eq:id}):
\[
=E[Y_1-Y_0|A=1]-E[Y_1-Y_0 |A=0]
\]

\end{document}